\documentstyle[preprint,aps,psfig]{revtex}

\newcommand {\rd}{\rho_{\scriptscriptstyle D}}
\newcommand {\cf}{{\rm \scriptscriptstyle CF}}
\newcommand {\kf}{k_{\scriptscriptstyle F}}
\newcommand {\vf}{v_{\scriptscriptstyle F}}

\begin{document}

\title {Coulomb drag in the quantum Hall $\nu = \frac{1}{2}$
      state: Role of disorder}

\author {Ady Stern and Iddo Ussishkin}
    
\address {Department of Condensed Matter Physics, The Weizmann
  Institute of Science, Rehovot 76100, Israel}

\date {July 24, 1997}

\maketitle

\begin {abstract}
  We consider Coulomb drag between two layers of two-dimensional
  electron gases subject to a strong magnetic field, with the Landau
  level filling factor in each layer being $\frac {1}{2}$. We find
  $\rd$ to be {\it very large}, as compared to the zero magnetic field
  case. We attribute this enhancement to the slow decay of density
  fluctuations in a strong magnetic field.  For a clean system, the
  linear $q$-dependence of the longitudinal conductivity,
  characteristic of the $\nu=1/2$ state, leads a unique temperature
  dependence---$\rd \propto T^{4/3}$. Within a semiclassical
  approximation, disorder leads to a decrease of the transresistivity
  as compared with the clean case, and a temperature dependence of
  $T^2\log T$ at low temperatures.
\end {abstract}

\pacs {}

\tightenlines

\section {Introduction}

In this paper we consider the transresistivity of a system of two
identical parallel two-dimensional electronic layers in close
proximity. A current $I_1$ is driven in one layer, while the current
in the other layer is kept zero.  Due to the interactions between the
electrons at the different layers, a voltage $V_2$ develops in the
second layer. For a square sample, the transresistivity is $\rd = -
V_2 / I_1$. First discussed twenty years ago \cite {Pog77},
transresistivity is a subject of recent study, both experimental and
theoretical \cite {GEM+91,SL91,ZM93,KO95}, and in particular when the
system is placed in a perpendicular magnetic field \cite
{HNL+96,SS94}.

Here, we consider the transresistivity in a strong magnetic field,
with the Landau-level filling fractions of the two layers being $\nu_1
= \nu_2 = \frac {1}{2}$. We have considered this problem in detail in
\cite {US97}, neglecting the role of disorder. Here we focus on the
latter. Due to length restrictions the discussion and references below
are rather brief. We refer the reader to our previous work for a more
elaborate discussion.

An experimental study of Coulomb drag in the regime of a partially
filled Landau level was recently carried out by Eisenstein, Pfeiffer
and West \cite{EPW97}. Two theoretical works \cite {KM96} considered
the problem from a composite-fermion approach.

\section {Coulomb drag between clean $\nu = \frac {1}{2}$ layers}
\label {sec:rev}

If the layers are sufficiently close, the main contribution to the
transresistivity comes from the Coulomb interactions between the
electrons. In the limit of weak interlayer coupling, both in the
absence and in the presence of magnetic field, the transresistivity is
given by \cite{ZM93}
\begin{equation}
  \label{donald}
  \rd = \frac{1}{8 \pi^2}
  \frac{h}{e^2}\frac{1}{T n^2}
  \int \frac{d{\bf q}}{(2\pi)^2}\int_0^\infty  
  \frac {\hbar \, d\omega}{\sinh^2 \frac {\hbar\omega}{2T}}
  \, q^2 \, [{\rm Im} \Pi ({\bf q},\omega)]^2 \,
  |U_{\rm sc}({\bf q},\omega)|^2 \, .
\end{equation}
Here $U_{\rm sc}$ is the screened interlayer Coulomb interaction,
$\Pi$ is the single layer density-density response function
(irreducible with respect to the Coulomb interaction), $n$ is the
average electron density in each layer and $T$ is the temperature.
Coulomb drag is a result of scattering events between electrons of
different layers, transferring momentum $\hbar {\bf q}$ and energy
$\hbar \omega$ between the layers.  Thus, it reflects the response of
the two layers at finite wavevector ${\bf q}$ and frequency $\omega$.

In addition to appearing explicitly in Eq.\ (\ref{donald}), the
response function $\Pi$ also determines the screening of the
interlayer interaction. Interaction is described in terms of $2 \times
2$ matrices (denoted by hats), with indices corresponding to the two
layers. The screened interlayer interaction $U_{\rm sc}$ is the
off-diagonal element of
\begin{equation}
  \label{screened}
  \hat V_{\rm sc} = \hat V_{\rm bare}
  \, (1 + \hat \Pi \hat V_{\rm bare})^{-1} \, ,
\end{equation}
where
\begin{equation}
  \label{matrixdef}
  \hat V_{\rm bare} = \frac {2 \pi e^2}{\epsilon q} \left(
    \begin{array}{ccc}
      1 &  & e^{-q d} \\
      e^{-q d} &  & 1
    \end{array}
  \right )  \; \; \; {\rm and} \; \; \; 
  \hat \Pi = \left (
    \begin{array}{ccc}
      \Pi & & 0\\
      0 & & \Pi
    \end{array}
  \right) \, .
\end{equation}

The transresistivity is determined by regions of ${\bf q}$ and
$\omega$ where the integrand of Eq.\ (\ref {donald}) is large. It is
thus interesting to look at the poles of the integrand. They are the
solutions of
\begin{equation}
  \label{pole}
  i \omega - \frac {q^2}{e^2} \sigma ({\bf q}, \omega)
  (V_b (q) \pm U_b (q)) = 0 \, ,
\end{equation}
where $\Pi$ was expressed in terms of the conductivity $\sigma$ for
physical clarification\cite{US97}. The solutions of (\ref {pole}) are the
dispersion relations for the decay of symmetric and antisymmetric
charge density modulations. In the limit of small $q$ they are given
by $i \omega \propto q \sigma (q,\omega)$ for the symmetric modulation, and
$i \omega \propto q^2 \sigma (q,\omega)$ for the antisymmetric modulation.

For $\nu=1/2$ the calculation of $\Pi$ is carried out in the
composite-fermion picture for the half-filled Landau level
\cite{HLR93}. Response functions are $2 \times 2$ matrices (denoted by
tildes), with entries corresponding to the density (denoted by $0$)
and to the transverse current (denoted by $1$). The electronic
response function $\tilde \Pi^e$ is expressed as
\begin{equation}
  \label{picf}
  ({\tilde\Pi}^{e})^{-1} = \left(
    \begin{array}{ccc}
      0 & & \frac {2 \pi i \hbar \tilde \phi}{q} \\
      -\frac{2 \pi i \hbar \tilde \phi}{q} & & 0
    \end{array} \right)
  + (\tilde \Pi^\cf)^{-1} \, .
\end{equation}
The matrix appearing explicitly in (\ref {picf}) is the Chern-Simons
interaction matrix, with $\tilde \phi = 2$ for $\nu = \frac {1}{2}$.
The composite-fermion response function $\tilde \Pi^\cf$ describes the
response of the composite fermions to the total scalar and vector
potentials, including external, Coulomb and Chern-Simons
contributions. The electronic density-density response function
appearing in Eqs. (\ref{donald})--(\ref{matrixdef}) is $\Pi =
\Pi^e_{00}$.

Within the random phase approximation, in the limit of $q \ll \kf$ and
$\omega \ll q \vf$, we have \cite {HLR93}
\begin{equation}
  \label{Pieq}
  \Pi ({\bf q}, \omega) = \frac {q^3}{
    q^3 \left( \frac {dn}{d\mu} \right)^{-1} -
    2 \pi i \hbar \tilde \phi^2 \omega \kf} \, ,
\end{equation}
where $dn/d\mu$ is the thermodynamical electronic compressibility of
the system.  This special form of $\Pi$ has several consequences.
First, it leads to the linear dependence on $q$ of the conductivity,
$\sigma(q) \propto q$. As a result, antisymmetric charge density
modulations have a very slow decay rate, with $i \omega \propto q^3$,
that eventually leads to a large transresistivity.  Second, the
integrand of Eq.\ (\ref{donald}) is dominated by $q$ and $\omega$
along the line $\omega \propto q^3$, limited by $\omega \approx T$.
Hence the important contribution comes from $q \approx q_0 (T) = \kf
(T / T_0)^{1/3}$, and not $q \approx d^{-1}$ as in the $B = 0$ case
($T_0$, typically $\approx 190^\circ$K, is defined below). Finally,
due to these considerations the temperature dependence of the
transresistivity is $\rd \propto T^{4/3}$. The leading temperature
dependence of $\rd$ is
\begin{equation}
  \label{rdt}
  \rd = 0.825 \frac {h}{e^2} \left( \frac {T}{T_0} \right)^{4/3} +
  {\cal O} (T^{5/3}) \, ,
\end{equation}
where
\begin{equation}
  \label{T0}
  T_0 = \frac {4 \pi e^2 n d}{\tilde \phi^2 \epsilon} 
  (1 + \alpha) \, , \; \; \; 
  \alpha = \frac {d n}{d \mu} \frac {2 \pi e^2 d}{\epsilon} \, .
\end{equation}
Typically $\alpha$ is a small parameter. The calculation sketched
above is described in detail in \cite{US97}.

\section {Semiclassical approximation for disorder}
\label {sec:dis}

The use of Eq.\ (\ref{Pieq}) is valid, in the presence of disorder,
for $q \gg l_{\rm el}^{-1}$, where $l_{\rm el}$ is the single-layer
composite fermion elastic mean free path. This puts a limit on the
validity of the calculations sketched above \cite {US97}, which may be
expressed in terms of the single layer longitudinal resistivity as
\begin{equation}
  \label{validity}
  \rho_{xx} \ll 2 \frac {h}{e^2}
  \left (\frac {T}{T_0} \right )^{1/3} \, .
\end{equation}
For typical experimental values this condition is barely fulfilled, as
$\rho_{xx} \approx 2500$~$\Omega$, while the right hand side of Eq.\ 
(\ref {validity}) is approximately 4000--8000~$\Omega$.

To include disorder, we first use the Boltzmann equation to find the
composite-fermion response function $\tilde \Pi^\cf$ \cite {SH93,MW}.
The Boltzmann equation for the composite fermion distribution function
$f ({\bf r}, {\bf k}, t)$ in the presence of an electric field $\bf E$
is given by
\begin{equation}
  \label{be}
  \frac {\partial f}{\partial t} +
  \frac {\partial f}{\partial {\bf r}} \cdot 
  \frac {\hbar {\bf k}}{m^*} -
  \frac {1}{\hbar} \frac {\partial f}{\partial {\bf k}} \cdot e {\bf E} = 
  \left( \frac {\partial f}{\partial t} \right)_{\rm sc} \, ,
\end{equation}
where $m^*$ is the composite fermion effective mass. With ${\bf E}
({\bf r}, t) = {\bf E} \exp (i {\bf q} \cdot {\bf r} - i \omega t)$,
one seeks a solution of the form
\begin{equation}
  \label{f}
  f ({\bf r}, {\bf k}, t) = f_0 (\epsilon_{\bf k}) +
  \left( - \frac {\partial f_0}{\partial \epsilon}
    (\epsilon_{\bf k}) \right) f (\phi)
  \exp (i {\bf q} \cdot {\bf r} - i \omega t)
  \, ,
\end{equation}
where $\phi$ is the polar angle in the ${\bf k}$ plane, and $f_0
(\epsilon)$ is the equilibrium Fermi-Dirac distribution function. The
scattering term is taken here in the relaxation time approximation
(see comments below)
\begin{equation}
  \label{dfdtsc}
  \left( \frac {\partial f (\phi)}{\partial t} \right)_{\rm sc} =
  - \frac {f (\phi) - \bar f}{\tau} \, ,
  \; \; \; {\rm where} \; \; \; \bar f = \frac {1}{2 \pi}
  \int_0^{2 \pi} d \phi \, f (\phi) \, .
\end{equation}
Here, $\tau$ is the composite-fermion transport mean free time. 

Solving Eq.\ (\ref{be}), $f$ is used to obtain the current in the
system, using
\begin{equation}
  {\bf j} ({\bf r}, t) = -e \int \frac {d {\bf k}}{(2 \pi)^2}
  \frac {\hbar {\bf k}}{m^*} f ({\bf r}, {\bf k}, t) \, ,
\end{equation}
from which the conductivity, and hence the response functions,
may be extracted. The result is
\begin{mathletters}
  \label{picfbe}
  \begin{eqnarray}
    \Pi^\cf_{00} & = & \frac {m^*}{2 \pi \hbar^2} \left( 1 +
      \frac {i \omega}{
        \sqrt {(1 / \tau - i \omega)^2 + (q \vf)^2} -
        1 / \tau} \right) \, ,
    \\
    \Pi^\cf_{11} & = & - \frac {q^2}{24 \pi m^*}
    - i \frac {\omega}{q^2}
    \frac {m^*}{2 \pi \hbar^2} \left( \frac {1}{\tau} - i \omega -
      \sqrt {(1 / \tau - i \omega)^2 + (q \vf)^2} \right) \, .
  \end{eqnarray}
\end{mathletters}
The diamagnetic term in $\Pi^\cf_{11}$ does not come out of
semiclassical calculation, and is added by hand. The electronic
response function $\Pi$ is found using (\ref {picf}). It has the
correct form, Eq.\ (\ref {Pieq}), in the clean limit ($\tau
\rightarrow \infty$). In the diffusive limit ($\omega \tau \ll 1$ and
$q l \ll 1$) we obtain the diffusive response function
\begin{equation}
  \label{pidis}
  \Pi = \frac {d n}{d \mu} \frac {D_e q^2}{-i \omega + D_e q^2} 
  \, , \; \; \;{\rm with}\ \ \ \ \ \ 
  D_e = \frac {1 + \tilde \phi^2 / 12}{1 + (\tilde \phi \kf l / 2)^2}
  \frac {\vf^2 \tau}{2} \, . 
\end{equation}
The simplest form of relaxation time approximation, Eq.
(\ref{dfdtsc}), may not be appropriate for the dynamics of composite
fermions \cite{HLR93,MW}. However, a diffusive motion is
described by a density--density response function of the form
(\ref{pidis}). Thus, we expect an improved relaxation time
approximation to affect the value of $D_e$, but not the functional
form of (\ref{pidis}).

For low enough temperatures ($q_0 (T) l \ll 1$) the transresistivity
is dominated by the diffusive regime, and is given by (cf.\ 
\cite{ZM93,KO95})
\begin{equation}
  \rd = \frac {2 \pi}{3} \frac {h}{e^2} 
  \left( \frac {T}{T_{\rm dif}} \right)^2
  \ln \left( \frac {T_{\rm dif}}{2 T} \right)
  \frac {1}{(\kf d)^4} \, , \; \; \; 
  T_{\rm dif} = \frac {4 \pi e^2}{\epsilon d} 
  \frac {dn}{d \mu} \frac {D_e}{\hbar}  \, .
\end{equation} 
Eq. (\ref{pidis}) may also describe a diffusive motion at zero
magnetic field, where the diffusion constant is usually much larger
than $D_e$. Indeed, at zero magnetic field and fantastically low
temperatures a $T^2\log{T}$ dependence of $\rd$ was predicted in
\cite{ZM93}. The smallness of the diffusion constant in a strong
magnetic field, as well as the emphasis on small $q$'s in the clean
limit, makes the $T^2\log{T}$ regime in the $\nu=1/2$ case hold at
experimentally accessible temperatures.

To obtain the transresistivity in the regime of interest,
interpolating the clean and disordered limit, the full expression
(\ref {picfbe}) for $\tilde \Pi^\cf$ must be used. Eq.\ (\ref {picf})
is then used to obtain $\Pi$, Eq.\ (\ref {screened}) to obtain $U_{\rm
  sc}$. The transresistivity is then given by Eq.\ (\ref {donald}).
The integration in Eq.\ (\ref {donald}) is carried out numerically.
Fig.\ \ref{fig:rd} shows the transresistivity for different values of
disorder. The parameters used in the calculation are $n = 1.4 \times
10^{11}$~${\rm cm}^{-2}$, $d = 200$~\AA, and $m = 4 m_b$ where $m_b$
is the bare mass of the electron in GaAs. The amount of disorder
included is moderate: At $T = 0.5^\circ$K we have $q_0 l = 8.0$,
$3.2$, and $1.6$ for $\rho_{xx} = 1000$~$\Omega$, $2500$~$\Omega$, and
$5000$~$\Omega$.  But, as can be seen in the figure, disorder reduces
the transresistivity rather significantly---by a factor of order 2.
The transresistivity remains much larger (by 3--4 orders of magnitude)
than at $B=0$.

\section {Summary}
\label {sec:sum}

In this paper we consider the Coulomb drag between two layers of
two-dimensional electron gases at Landau level filling fractions of
$\nu_1 = \nu_2 = \frac {1}{2}$. The coupling between the layers is
discussed in purely electronic terms.  We find
\begin{itemize}
\item Coulomb drag at $\nu=1/2$ is {\it much larger} than in the
  absence of a magnetic field. The enhancement is due to the slow
  decay of density fluctuations in a strong magnetic field.
\item For the clean case, we find that a unique temperature dependence
  of the transresistivity---$\rd \propto T^{4/3}$.
\item At $\nu=1/2$ disorder reduces Coulomb drag (as opposed to its
  role in the absence of a magnetic field), leading to a $T^2\log{T}$
  dependence at low temperatures.  In typical experimental values the
  effect of disorder on $\rd$ is not negligible.
\end{itemize}
We thank J.~P. Eisenstein and B.~I. Halperin for useful discussions.
This research was supported by the US-Israel Binational Science
Foundation (95--250/1) and by the Minerva foundation (Munich,
Germany). A.~S. is supported by the V. Ehrlich career development
chair.

\setlength{\unitlength}{8.4cm}

\begin{figure}
  \begin {center}
    \vspace {1cm}
    \begin{picture}(1,0.618)
      \put (0,0){\psfig {figure=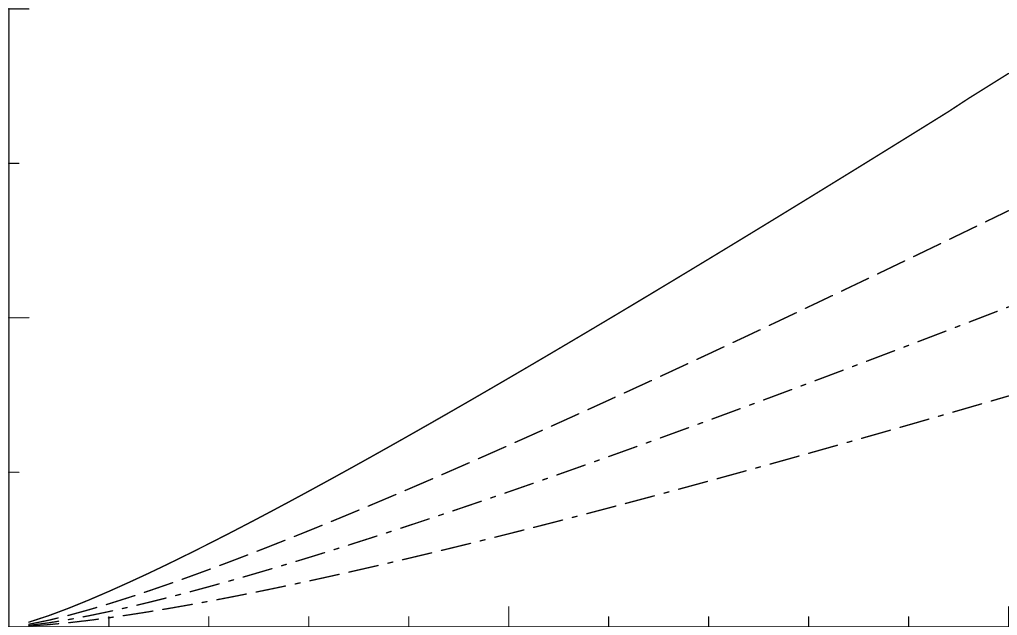,width=8.4cm}}
      \put (0,0.668){\makebox(0,0)[b]{
          $\rho_{\scriptscriptstyle D} [\Omega]$}}
      \put (1.02,0){\makebox(0,0)[l]{$T [^\circ {\rm K}]$}}
      \put (1,-0.01){\makebox(0,0)[t]{$1$}}
      \put (0.5,-0.01){\makebox(0,0)[t]{$0.5$}}
      \put (-0.01,0.618){\makebox(0,0)[r]{$20$}}
      \put (-0.01,0.309){\makebox(0,0)[r]{$10$}}
      \put (1.02,0.618){\makebox(0,0)[bl]{$\rho_{xx}$}}
      \put (1.02,0.553){\makebox(0,0)[l]{$0 \Omega$}}
      \put (1.02,0.416){\makebox(0,0)[l]{$1000 \Omega$}}
      \put (1.02,0.32){\makebox(0,0)[l]{$2500 \Omega$}}
      \put (1.02,0.231){\makebox(0,0)[l]{$5000 \Omega$}}
    \end{picture}
    \vspace {1cm}
  \end {center}
  \caption{The transresistivity as a function of temperature,
    for different values of disorder.  Details are given in the text. }
  \label{fig:rd}
\end{figure}

\end{document}